\documentclass[english,prl,aps,twocolumn,10pt]{revtex4-2}

\usepackage{amsmath}
\usepackage{amssymb}
\usepackage{graphicx}
\usepackage{babel}
\usepackage{array}
\usepackage{verbatim}
\usepackage[colorlinks=true, pdfstartview=FitV, linkcolor=blue, citecolor=blue, urlcolor=blue]{hyperref} % enable links

\usepackage{subfigure}

\def\NVm	  {{NV$^-$}}
\def\NVo	  {{NV$^0$}}

\newcommand{\mr}[1]{\mathrm{#1}}
\newcommand{\unit}[1]{\,\mathrm{#1}}

\newcommand{\us}{\,\mu{\rm s}}

\newcommand{\uW}{\,\mu{\rm W}}

\newcommand{\ket}[1]{\ensuremath{\left|#1\right\rangle}}
\newcommand{\braket}[2]{\ensuremath{\left\langle#1|#2\right\rangle}}
\newcommand{\up}{\ket{\uparrow}}
\newcommand{\down}{\ket{\downarrow}}

\newcommand{\fa}{f_\mr{a}}

\newcommand{\fbi}{f_{\mr{b-d}}}
\newcommand{\Mup}{M_\uparrow}
\newcommand{\Mdown}{M_\downarrow}
\newcommand{\ms}{m_S}
\newcommand{\Np}{\mathcal{N}_p}
\newcommand{\Nup}{\mathcal{N}_\uparrow}
\newcommand{\Nupdown}{\mathcal{N}_{\uparrow,\downarrow}}
\newcommand{\Ndown}{\mathcal{N}_\downarrow}
\newcommand{\NVa}{NV$_\text{a}$}
\newcommand{\NVbi}{NV$_{\text{b-d}}$}
\newcommand{\NVall}{NV$_{\text{a-d}}$}

\newcommand{\tpi}{t_\pi}
\newcommand{\tRO}{t_\mathrm{RO}}
\newcommand{\tOS}{t_\mathrm{OS}}
\newcommand{\Vup}{V_\uparrow}
\newcommand{\Vdown}{V_\downarrow}
\newcommand{\Vclass}{\widetilde{V}_\mr{RO}}
\newcommand{\Vproj}{\widetilde{V}_\mr{proj}}
\newcommand{\Vtot}{\widetilde{V}}

\begin{document}

\title{Spin counting via projection noise measurement \\ of mesoscopic solid-state spin ensemble}

\author{L.~Bechelli$^{1}$, K.~Herb$^{1,\ddagger}$, L.~A.~V\"olker$^{1}$, and C.~L.~Degen$^{1,2}$}
\email{degenc@ethz.ch}
\altaffiliation{$^\ddagger$Present address: JILA, National Institute of Standards and Technology and the University of Colorado, Boulder, Colorado 80309-0440, USA.}
\affiliation{$^1$Department of Physics, ETH Zurich, Otto Stern Weg 1, 8093 Zurich, Switzerland.}
\affiliation{$^2$Quantum Center, ETH Z\"urich, 8093 Z\"urich, Switzerland.}

\begin{abstract}
Quantum projection noise is the fundamental noise source for the population measurement of spin ensembles. While projection-noise-limited measurements have been extensively studied in atomic systems, corresponding experiments on solid-state spin ensembles remain challenging due to dominant classical readout noise. Here, we report direct measurement of the quantum projection noise of mesoscopic ensembles of nitrogen-vacancy (NV) spin defects at room temperature.  Our experiment is enabled by a high optically-detected magnetic resonance (ODMR) contrast of over 20\% for a single crystallographic orientation of the defect spins, obtained by combining polarization-selective optical excitation with spin-to-charge conversion.  We use our protocol to demonstrate projection noise measurements and spin counting from nanoscale NV ensembles of up to 43 spins.  We further demonstrate that the protocol allows for significant gains in sensitivity for magnetometry applications without need for cryogenic operation or high bias magnetic fields
\end{abstract}
	
\date{\today}
	
\maketitle

%%% Introduction

\begin{figure}
	\centering
	\includegraphics[width=0.98\columnwidth]{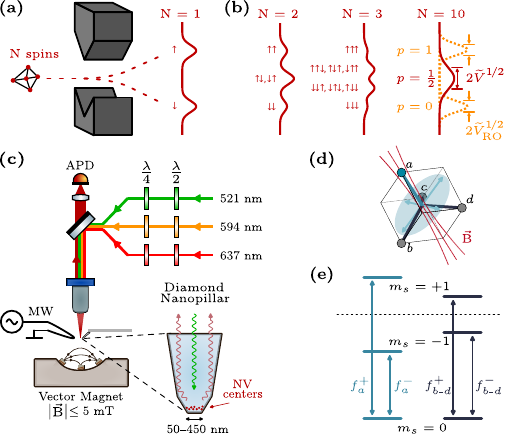}
	\caption{Concept of the spin counting experiment.
		(a) Stern-Gerlach-type experiment with a rigid cluster of $N$ independent spins-1/2 particles. Projective measurement along an axis orthogonal to the spins' initial direction randomly collapses each spin onto either $\up$ and $\down$.
		(b) For $N>1$ the projection assumes a binomial distribution with variance $\Vproj$.  The variance is largest at maximum superposition ($p=0.5$) and zero for eigenstates ($p=0,1$). Classical readout noise leads to a broadening with variance $\Vclass$.  By measuring $\Vproj$, $N$ can be inferred.
		(c) Experimental setup.  A confocal fluorescence microscope with three-color laser excitation and polarization control ($\lambda/2$ and $\lambda/4$ plates) is used together with a single photon detector (APD) for controlling and detecting the spin and charge states of NV centers.  Microwave (MW) fields are used for coherent spin manipulation and an external magnetic field $\vec{B}$ is used to split the spin resonances. Inset: the sample is a diamond nanopillar containing $\sim 10^1-10^3$ NV centers at the apex.
		(d) Crystallographic orientations of the four NV center families, \NVall.  We align the external field $\vec{B}$ (red) with \NVa\ (light blue), leaving the other three orientations \NVbi\ (dark blue) misaligned.
		(e) Spin energy levels $\ms=0,\pm1$ and transitions frequencies $\fa^\pm$, $\fbi^\pm$ for \NVa\ and \NVbi, respectively.
	}
	\label{fig1}
\end{figure}

The intrinsically probabilistic character of quantum measurements has been a defining feature of quantum physics since its inception. In quantum mechanics, the act of measurement projects a system onto an eigenstate of the measured observable, and the recorded outcome corresponds to the associated eigenvalue.  For the canonical example of a single spin-1/2 particle prepared in a superposition state, a measurement randomly collapses the state onto one of two possible eigenstates, giving rise to quantum projection noise~\cite{itano1993}~(Fig.~\ref{fig1}a).  The quantum projection noise defines the standard quantum limit (SQL) of independent particles~\cite{caves1981} and imposes a fundamental bound on the precision with which a quantum state can be inferred.

For a cluster of $N$ identical and uncorrelated spins, the collective observable admits $N+1$ distinct projections (Fig.~\ref{fig1}b). As $N$ becomes large, the distribution of possible outcomes becomes increasingly fine-grained and eventually approaches a quasi-continuous form. As a result, the individual projections are challenging to resolve experimentally.
Nevertheless, in systems with a highly-controlled environment, such as cold atomic gases, projection noise in ensembles can be directly observed and investigated~\cite{santarelli1999,koschorreck2010}.  These systems have further enabled the generation of spin squeezing and entangled states that reduced quantum noise below the SQL~\cite{hosten2016,appel2009,hald1999,yang2025} and established projection noise as a standard reference in quantum metrology~\cite{yurke1986,giovannetti2006}.

By contrast, detecting the projection noise of solid-state spin ensembles is much more demanding owing to the heterogeneous and noisy environment of the solid host. Although the random polarization fluctuations of electronic or nuclear spins have been measured~\cite{mamin2003,degen2007}, these are uncontrolled and resemble shot noise rather than quantum projection noise.
Taking the prototypical example of the nitrogen-vacancy (NV) center in diamond~\cite{doherty2013}, charge-state instabilities, strain gradients, inhomogeneous broadening, imperfect spin control, and averaging over multiple crystallographic orientations hamper efficient detection of the spin state.  These issues are shared to varying degree with other prospective spin platforms~\cite{koehl2011,rizzato2023}.
As a result, the technical readout noise in the ensembles generally far exceeds the fundamental projection noise limit, even when optimized readout protocols are used~\cite{jayakumar2018,hopper2018}. While projection noise has been experimentally resolved for a single NV center~\cite{shields2015,zhao2024sciadv} and recent experiments have demonstrated spin squeezing and spin amplification in NV ensembles~\cite{wu2025,gao2025}, experiments aimed at measuring the projection noise have only been considered recently~\cite{maier2025}. 

%%% Here, we...

Here, we report measurements of the quantum projection noise and spin counting in mesoscopic ensembles of NV centers in diamond at room temperature.
First, we demonstrate that a combination of polarization-selective optical excitation~\cite{magaletti2024} and spin-to-charge conversion (SCC)~\cite{shields2015} allows for an improved spin readout with an ODMR contrast of over 20\% for a single crystallographic orientation of the NV ensemble.
Next, we use the improved readout to realize measurement of the quantum projection noise from up to $N\sim 40$ spins.  Thereby, the variance of the projection noise provides a calibration-free and model-independent count of the number of spins contributing to the signal.
Our approach, demonstrated on nanoscale diamond tips, neither requires cryogenic conditions nor high bias magnetic fields and is well-suited for advancing applications in sensitive magnetometry and many-body spin physics.

%%% Theory

%\section*{Theory}

For our experiment, we consider an ensemble of $N$ identical spin-1/2 particles in a rigid cluster. When spins prepared in a superposition state $\ket{\psi} = a_0\up + a_1\down$, where $a_0$, $a_1$ are complex coefficients with $|a_0|^2+|a_1|^2=1$, are subjected to a projective measurement, the state will collapse onto one of the two eigenstates $\up$ and $\down$ with probabilities $\left\vert\braket{\uparrow}{\psi}\right\vert^2=|a_0|^2=p$ and $\left\vert\braket{\downarrow}{\psi}\right\vert^2=|a_1|^2=1-p$, respectively.  Unless the spins are entangled, each of the $N$ spins projects independently, leading to a binomial distribution as shown in Fig.~\ref{fig1}b.  The variance of this binomial distribution is given by $\sigma^2 = Np(1-p)$~\cite{lehmann1998}.  Clearly, $\sigma^2$ is minimized for basis states ($p=0$ and $p=1$) while it is maximized at equal superposition ($p=0.5$), see Fig.~\ref{fig1}b.

The variance $\sigma^2$ is not directly accessible experimentally, but must be mapped onto a measurable physical quantity~\cite{degen2017}.  For optical detection, considered in our experiment, the measured quantity is the number of photons $c$ reflecting an optical intensity.  The measurement of $c$ is itself a statistical process that adds extra noise, which we refer to as (classical) readout noise.  In general, the measurement process is described by different probability distributions $\Nup(c)$ and $\Ndown(c)$ of the number of photons $c$ collected conditional on the spin states $\up$ and $\down$, respectively.  Below, we find that the $\Nupdown(c)$ are described by negative binomial (Pascal) distributions.  Regardless of the details of $\Nupdown(c)$, the mean and variance of $c$ are well-defined quantities, given by the weighted averages of the two distributions~\cite{supplemental}.  As a result, the total variance of the measured photon count of $N$ independent spins is given by
\begin{align}
	V(p)
	%&= \Vproj(p) + \Vclass(p) \\
	&= \frac{1}{N} p(1-p) (\Mup-\Mdown)^2 + \left[ p\Vup + (1-p)\Vdown \right]
	\label{eq:variance_total_unnormalized}
\end{align}
where $\Mup$,$\Mdown$ are the means and $\Vup$,$\Vdown$ the variances of photon counts measured when the spins are initialized in $\up$,$\down$ respectively~\cite{supplemental}.

Dividing Eq.~(\ref{eq:variance_total_unnormalized}) by $(\Mup-\Mdown)^2/4$, we obtain the measurement noise variance normalized to the projection noise of a single spin,
\begin{align}
	\Vtot(p)
	%= \frac{V(p)}{V_0}
	%&= \Vproj(p) + \Vclass(p) \\
	&= \frac{4p(1-p)}{N}  + \frac{4[p\Vup + (1-p)\Vdown]}{(\Mup-\Mdown)^2} .
	\label{eq:variance_total}
\end{align}
In the context of quantum sensing, the square root of $\Vtot$ is directly proportional to the sensitivity and takes its scaling with the number of spins into account~\cite{barry2020}.  In Eq.~(\ref{eq:variance_total}), the first term is the projection noise variance, $\Vproj=4\sigma^2/N^2=4p(1-p)/N$, and the second term is the readout noise variance, $\Vclass=4[p\Vup + (1-p)\Vdown]/(\Mup-\Mdown)^2$.
As expected, $\Vproj$ scales as $1/N$ with increasing ensemble size $N$, reflecting the statistical averaging over independent spins. We note that this is consistent with the SQL scaling of $1/\sqrt{N}$ of the sensitivity typically reported in the literature~\cite{santarelli1999, giovannetti2006}.   $\Vproj$ is a fundamental noise contribution that cannot be reduced further without using resources such as spin squeezing~\cite{wu2025}.
By contrast, $\Vclass$ is inversely proportional to the number of detected photons and can be reduced to an arbitrarily low level using averaging, as long as the readout is dominated by photon shot noise ($\Vup\approx\Mup$ and $\Vdown\approx\Mdown$).
Because all quantities in Eq.~(\ref{eq:variance_total}) (beside $N$) can be determined by measuring the means and variances of photon counts for initial states $\up$ and $\down$, the number of spins $N$ can be directly extracted from $\Vtot(p)$ without further assumptions on the distributions $\Nupdown(c)$.

%%% Experimental
%\section*{Experimental}

\begin{figure}
	\centering
	\includegraphics[width=0.99\columnwidth]{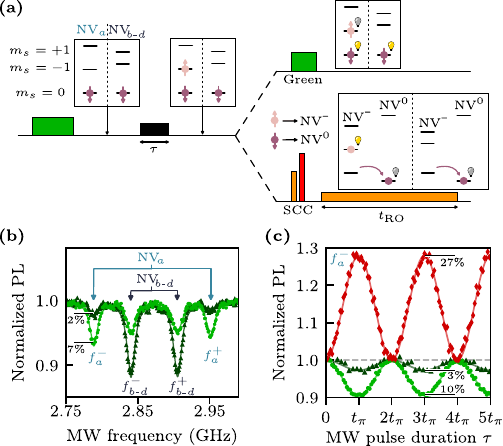}
	\caption{Orientation-selective excitation and readout.
		(a) Schematic of the sensing and readout protocols. After green initialization of charge and spin states for all orientations, selective microwave manipulation is applied to the aligned NV orientation. Readout is performed either \textit{via} standard green fluorescence detection (upper path) or \textit{via} charge state detection using SCC (lower path). Boxes show the population of energy levels for \NVa\ and \NVbi.
		(b) Pulsed ODMR spectra acquired using standard green readout without polarization control (dark green triangles) and for optimized polarization (light green dots) of the green laser pulse. PL, photoluminescence.
		(c) Rabi oscillations of the $\fa^-$ transition measured using standard green readout with no polarization control (dark green triangles), optimized polarization (light green dots) and for SCC readout (red diamonds). The $\pi$-pulse duration is $\tpi\sim 40\unit{ns}$.
		}
	\label{fig2}
\end{figure}

We experimentally demonstrate spin counting using an ensemble of NV centers in a diamond nanopillar tip~\cite{babinec2010,momenzadeh2015,zhu2023} created by shallow ion implantation.  The tips have a multicone structure~\cite{zhu2023} with top diameters ranging from approximately $50$ to $450\unit{nm}$ corresponding to between $10^1-10^3$ NV centers per tip~\cite{bechelli2026multi}.
The experimental setup, shown in Fig.~\ref{fig1}c and detailed Ref.~\cite{supplemental}, consists of a home-built confocal microscope operated at room temperature.  The optical excitation includes three laser sources at wavelengths of $521\unit{nm}$ (green), $594\unit{nm}$ (orange), and $637\unit{nm}$ (red) to allow for charge-state control~\cite{shields2015,bluvstein2019}, each with its own polarization control. 
Experiments are performed in an external bias field of $\sim 5\unit{mT}$ that is aligned with one of the four crystallographic NV orientations (\NVa, see Fig.~\ref{fig1}d), the other three orientations (\NVbi) being misaligned by $\sim 70^\circ$.  In this configuration, the spin resonance spectrum displays four peaks, reflecting two spin transitions ($\ms=0\leftrightarrow\pm1$) and two sub-ensembles (\NVa, \NVbi). The associated energy level diagram and transition frequencies $\fa^\pm$ and $\fbi^\pm$ are depicted in Fig.~\ref{fig1}e, and a representative ODMR spectrum is shown in Fig.~\ref{fig2}b.

%%% Results: Orientation selection
%\section*{Results}
%\subsection*{Orientation-selective spin detection}

\textit{Orientation-selective spin detection -- }
In a first step, we show that a combination of polarization-selective optical excitation~\cite{epstein2005,magaletti2024} and SCC~\cite{shields2015} allows for large improvements of the ODMR contrast.  To implement polarization-selective excitation, we adjust the polarization of the green laser (Fig.~\ref{fig1}d) to maximize the overlap with optical dipoles of the \NVa\ orientation, thus enhancing the excitation of the \NVa\ subensemble while reducing the excitation from the \NVbi.
In Fig.~\ref{fig2}b we show pulsed ODMR spectra~\cite{dreau2011} without (dark green) and with (light green) optimization of the green laser polarization.  We find that the ODMR contrast increases from 2\% without polarization control, which is typical for NV ensembles~\cite{tetienne2018,kageura2024,noda2025}, to approximately 7\% with polarization control, corresponding to an improvement by over $3\times$.  At the same time, the intensity of the \NVbi\ resonances ($\fbi^\pm$) is reduced by over a factor of two.  Fig.~\ref{fig2}b confirms that the polarization of the exciting laser pulse has a strong impact on the readout contrast in ensembles.  Nevertheless, the orientation selection is not perfect because the \NVbi\ optical dipoles still partially overlap with the green laser polarization.

\begin{figure*}[tb!]
	\centering
	\includegraphics[width=0.99\textwidth]{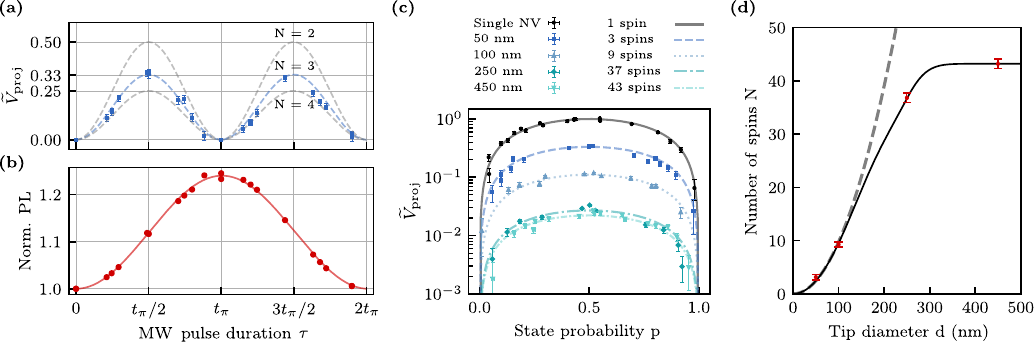}
	\caption{Spin counting experiment.
		(a) Projection noise variance as a function of Rabi angle $\phi = \pi\tau/\tpi$ (blue dots) together with the expected projection noise for discrete numbers of spins $N$ (gray).
        (b) Corresponding normalized PL showing the Rabi oscillation of the spin ensemble.
		(c) Projection noise variance measured for diamond tips with different apex diameters $d$ (blue shades) as well as for a single NV center (black). For each dataset, the projection noise is fitted with $N$ as the only free parameter (curves).
		(d) Number of spins $N$ as a function of tip diameter $d$. The gray dashed line is an estimate for the number of \NVa\ spins based on the implantation density and a conversion efficiency of 5\%~\cite{bechelli2026multi}.  The black solid line takes the finite mode volume of the optical excitation into account~\cite{supplemental}.
		Error bars are one standard deviation~\cite{supplemental}.
	}
	\label{fig3}
\end{figure*}

To further suppress the response from the \NVbi, we use the SCC scheme~\cite{shields2015} (Fig.~\ref{fig2}a) to selectively convert these orientations to the neutral \NVo\ charge state.  Because \NVo\ is not excited by the subsequent orange readout pulse~\cite{doherty2013,shields2015}, the emitted photons solely report on the remaining \NVm\ population. 
Fig.~\ref{fig2}c demonstrates the combined improvement of polarization control and SCC readout by recording Rabi oscillations of the \NVa\ sub-ensemble. Using standard green readout with non-optimized polarization as a benchmark, the Rabi contrast is $\sim 3\%$ (dark green) consistent with the result of Fig.~\ref{fig2}b. Optimizing polarization increases the contrast to $\sim 10\%$ (light green). Further adding SCC orientation selection, the contrast improves to $\sim 27\%$ for this tip (23\% when normalized to the peak PL), reaching values close to those of perfectly aligned ensembles~\cite{ishiwata2017} and single NV centers~\cite{bechelli2026multi}.  Measurements on other tips show contrast figures between $18-23\%$~\cite{supplemental}.  The data of Fig.~\ref{fig2}c constitutes a more than sevenfold improvement in contrast compared to the standard green readout without polarization control, and a twofold improvement compared to previous SCC measurements on NV ensembles~\cite{jayakumar2018,hopper2018}.

%%% Results: Spin counting
%\subsection*{Spin counting}

\textit{Spin counting -- }
Next, we demonstrate that the enhanced readout contrast afforded by the orientation selection protocol enables counting the number of spins $N$ in the ensemble. For this purpose, we prepare spins in a controlled superposition (Fig.~\ref{fig1}b) by driving Rabi oscillations with varying rotation angles $\phi$ on the $\fa^-$ transition of the \NVa\ family.  For each $\phi$ value, we repeat the sequence [Fig.~\ref{fig2}a] many times and store the number of photons for each repetition.  The resulting record of photon counts then allows for calculating the photon distribution $\Np(c)$, variance $\Vtot(p)$ and mean $M(p)=p\Mup + (1-p)\Mdown$, where $p=\sin^2(\phi/2)$~\cite{degen2017}.  In particular, from $\phi=0$ ($p=0$) we obtain $\Vup$, $\Mup$, and from $\phi=\pi$ ($p=1$) we obtain $\Vdown$, $\Mdown$, respectively.
Armed with these values, we determine the projection noise as
\begin{align}
	\Vproj(p) = \Vtot(p) - \Vclass(p)
	\label{eq:projection}
\end{align}
where $\Vtot(p)$ is the experimental measurement and where $\Vclass(p)$ is determined by Eq.~(\ref{eq:variance_total}). 

The resulting projection noise variance $\Vproj$ is shown in Fig.~\ref{fig3}a as a function of $\phi$.  As expected, the projection noise oscillates as $\sin^2\phi$ and reaches a minimum when the system is prepared in one of the two basis states ($p=0,1$ corresponding to $\up$,$\down$), as well as a maximum when the spins are initialized in an equal superposition ($p=0.5$). Importantly, the projection noise variance is bounded by $1/N$ [see Eq.~(\ref{eq:variance_total})], where $N$ is the number of independent spins that contribute to the measurement. Because $N$ is discrete and assuming all spins contribute equally to the signal (discussed below), we can therefore directly read-off the number of spins ($N=3$) from the variance measurement in Fig.~\ref{fig3}a.  Note that the estimate of $N$ is independent of the photon distribution $\mathcal{N}(c)$ and fully determined by the mean and variance of photon counts, $M(\phi)$ and $V(\phi)$.

\begin{figure}[tb!]
	\centering
	\includegraphics[width=0.99\columnwidth]{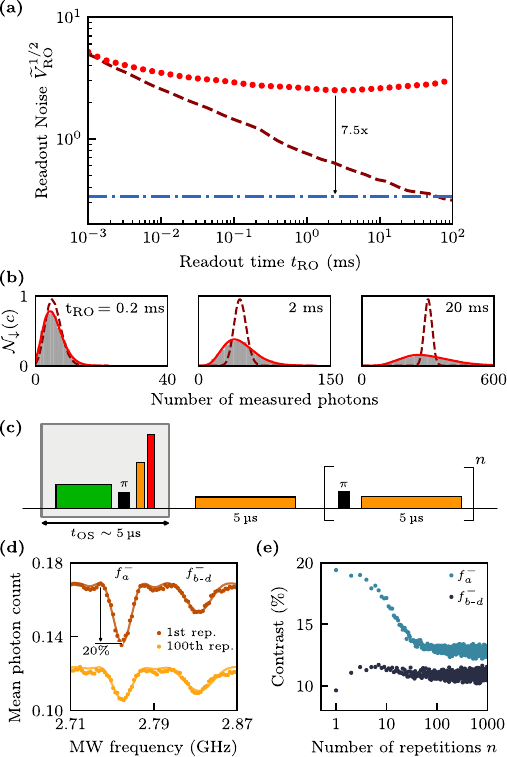}
	\caption{Standard quantum limit and contrast-enhanced magnetometry.
		(a) Readout noise $\Vclass^{1/2}$ as a function of orange readout time $\tRO$ [cf.~Fig.~\ref{fig2}a] for an ensemble of $N=9$ spins.  Dots reflect the experimental data.  The dash-dotted line is the SQL ($\Vproj$ for $N=9$) and the dashed line is the Poissonian limit.
		(b) Measured photon-count distributions (gray) for three representative $\tRO$. Solid curves are fits to a negative binomial distribution and dashed curves are the expected Poisson distribution.
		(c) Pulse sequence for orientation selection (gray box) followed by an arbitrary pulse sequence, here the measurement of a pulsed ODMR spectrum. 
		$n$ is the number of repetitions before orientation selection is refreshed.
		(d) Pulsed ODMR spectra for $n=1$ and $n=100$, using $\tRO=5\unit{\us}$ and $P=6\unit{\uW}$. 
        (e) ODMR contrast of the selected orientation ($\fa^-$ and $\fbi^-$) as a function of $n$.
	}
	\label{fig4}
\end{figure}

The measurement of Fig.~\ref{fig3}a is performed with the smallest tip diameter ($d\sim 50\unit{nm}$) with approximately 10 NV centers per tip, consistent with $\sim 2-3$ NV centers per orientation. In Fig.~\ref{fig3}c, we demonstrate spin counting measurements for tips with larger apex diameters $d$ of up to $450\unit{nm}$ hosting ensembles with up to $\sim 10^3$ NV centers~\cite{bechelli2026multi}.  We find that for all tips, the data are well described by $\Vproj\propto\sin^2\phi$ with the number of spins $N$ as the only free parameter.  We also observe that the ensemble size saturates around $N\sim 43$ for large $d$, which we attribute to the limited mode volume of the optical excitation.  As $d$ increases beyond $\sim 250\unit{nm}$, the number of NV centers contributing to the signal becomes limited by the laser excitation volume rather than the physical size of the tip~\cite{bechelli2026multi}.  In this limit, $N$ represents an effective number of spins that takes the relative contributions of each spin to $\Vproj$ into account~\cite{supplemental}.

%%% Results: Projection vs. readout noise
%\subsection*{SQL}

\textit{Standard quantum limit (SQL) -- }
Opposite to a recent experiment employing repetitive nuclear-assisted spin readout~\cite{maier2025}, our orientation-selection protocol does not yet operate at the projection noise limit.  To estimate how closely our experiment can approach the SQL, we prepare the spins in one of the basis states ($\downarrow$) where $\Vtot=\Vclass$ and compare it to $\Vproj(p=0.5)$.  Fig.~\ref{fig4}a plots $\Vclass(\tRO)$ as a function of orange readout time $\tRO$ (see Fig.~\ref{fig2}a).  We observe that while the readout noise initially decreases as $\Vclass\propto\tRO^{-1}$ as expected from photon shot-noise statistics, $\Vclass$ reaches a minimum between $\tRO\sim 0.1-10\unit{ms}$ and even increases for longer duration.  At the minimum, $\Vclass$ is approximately $7\times$ above the SQL.

The deviation from shot-noise statistics for longer $\tRO$ is well known and attributed to inadvertent charge-state conversion between \NVm\ and \NVo\ during optical illumination~\cite{beha2012,aslam2013,hacquebard2018}.  The effect is enhanced for near-surface NV centers~\cite{bluvstein2019,irber2021,mahdia2026}.  As a consequence, for long $\tRO$, the photon distribution changes from Poissonian to a broad negative binomial distribution~\cite{shields2015} (Fig.~\ref{fig4}b).  Therefore, the time scale of charge-state dynamics during the readout limits the maximum $\tRO$, and therefore, the minimum $\Vclass$ that can be reached in our orientation selection protocol.  Looking forward, we note that this limitation is technical rather than fundamental.  Demonstrated mitigation strategies include charge stabilization using surface functionalization~\cite{giri2023} or co-doping~\cite{geng2023} and optimized initialization~\cite{mahdia2026,song2020,wirtitsch2023} and readout schemes~\cite{hopper2016,zhao2024sciadv}.  Given that our present experiment is within an order of magnitude of the SQL, there is reasonable prospect that the SQL can be reached in optimized experiments.

%%% Results: Optimized magnetometry
%\subsection*{Optimized magnetometry}

\textit{Optimized magnetometry -- }
Finally, we demonstrate that even without operating at the SQL, our protocol has the potential to greatly improve the sensitivity of ensemble NV experiments.  Presented in Fig.~\ref{fig4}c, we insert  the orientation selection as a generic building block before an arbitrary spin manipulation sequence.  Here, the purpose of the orientation selection is to ``switch off'' the \NVbi\ by converting them to \NVo, while leaving \NVa\ in \NVm. Hence, the subsequent spin manipulation and readout steps only act on the \NVa\ sub-ensemble without unwanted interference from the \NVbi.  
In Fig.~\ref{fig4}d, as an example, we show ODMR spectra taken after the orientation selection, reaching a contrast of $\sim 20\%$ on the \NVa\ sub-ensemble.  Notably, the orientation-selection sequence is fast (here $\tOS \sim 5\unit{\us}$), adding minimum timing overhead.  Further, because the charge states are relatively robust under orange excitation, many repetitions of the sensing sequence can be executed before the orientation selection needs to be refreshed.  For the example in Fig.~\ref{fig4}e, the contrast remains high up to $n\approx 10$ repetitions.

%%% Summary and Outlook
%\section*{Discussion/Outlook}

\textit{Outlook -- }
Our efficient readout near the projection limit will greatly aid the study of spin dynamics in dense spin ensembles.  These form a promising platform for the study of many-body spin systems~\cite{choi2020}, dynamics of spin clusters~\cite{rosenfeld2018}, critical thermalization~\cite{kucsko2018}, noise characterization and study of noise models~\cite{Ajoy2019}, covariance measurements~\cite{rovny2024} as well as spin squeezing and amplification~\cite{wu2025,gao2025}. 
The high readout contrast of $>20\%$ will also directly improve the sensitivity for various uses in ensemble magnetometry, including vector sensing~\cite{broadway2018ne,broadway2020,barry2020}, scanning probe imaging~\cite{simon2021,bechelli2026multi}, multiplexed sensing~\cite{shim2022,huxter2025}, circuit imaging~\cite{garsi2024}, thermometry and magnetometry in biological samples~\cite{lesage2013, kucsko2013}, and inertial navigation~\cite{ledbetter2012}.
Importantly, our approach solely relies on the control of the optical excitation with minimal timing overhead and neither requires cryogenic conditions~\cite{robledo2011nature} nor high magnetic bias fields~\cite{jiang2009,maier2025}, and is therefore compatible with a wide range of measurement scenarios.

%%%%%%%%%%%%%% Acknowledgments

\textit{Acknowledgements -- }
The authors thank J.~Abendroth, T.~H\"achler, K.~Kim and P.~Perrin for helpful discussions and assistance with the setup, and P.~London, A.~Bleszynski-Jayich, M.~Cambria and S.~Kolkowitz for advice on spin-to-charge conversion.
This work has been supported by Swiss National Science Foundation (SNSF) through Grant No. 200020\_212051/1 and No. CRSII-222812, and by the State Secretariat for Education, Research, and Innovation (SERI) though Grant No. UeM019-8, 215927.

%%%%%%%%%%%%%% Bibliography
	
%\bibliography{C:/ETH/labview/library/library}

\end{document}